\documentclass{aastex63}

\shortauthors{Palit et al.}
\begin{document}
%\title{Template \aastex Article with Examples: v6.3\footnote{Released on June, 10th, 2019}}
\title{Clumpy wind accretion in Cygnus X-1}
\correspondingauthor{Ishika Palit}
\email{palit@cft.edu.pl}
\author[0000-0002-7827-4517]{Ishika Palit}
\affiliation{Center for Theoretical Physics PAS,
Al. Lotnikow 32/46,
02-668 Warsaw,
Poland}
\author[0000-0002-1622-3036]{Agnieszka Janiuk}
\affiliation{Center for Theoretical Physics PAS,
Al. Lotnikow 32/46,
02-668 Warsaw,
Poland}
\author[0000-0001-5848-4333]{Bozena Czerny}
\affiliation{Center for Theoretical Physics PAS,
Al. Lotnikow 32/46,
02-668 Warsaw,
Poland}
%%%%%%%%%%%%%%%%%%%%%%%%%%%%%%%%%%%%%%%%%%%%%%%%%%%%%%%%%%%%%%%%%%%%%%%%%%%%%%%%%%%%%%%%%%%%%%%

\begin{abstract}
Cygnus X-1 is one of the brightest X-ray sources observed and shows the X-ray intensity variations on time scales from milliseconds to months in both the soft and hard X-rays. The accretion onto the black hole is believed to be wind fed due to focused stellar wind from the binary companion HDE-226868. We aim to understand the physical mechanism responsible for the short timescale X-ray variability ($<$100 s) of the source in its Hard/Low state.
%accretion rate that is related to the variability of the X-ray flux from this source.
We compute the 2D relativistic hydrodynamic simulation of the low angular momentum accretion flow with a time dependent outer boundary condition that reflects the focused, clumpy wind from the super-giant in this X-ray binary system.
We follow the dynamical evolution of our model for about 100 s and present the results showing an oscillatory shock, being a potential explanation of variability observed in hard X-rays. The simulated model with shock solutions is in good agreement with the observed power density spectra of the source. 
\end{abstract}

\keywords{accretion, accretion disks, black hole physics, X-ray binaries, Cygnus X-1, stellar winds, gravitation, hydrodynamics}

%%%%%%%%%%%%%%%%%%%%%%%%%%%%%%%%%%%%%%%%%%%%%%%%%%%%%%%%%%%%%%%%%%%%%%%%%%%%%%%%%%%%%%%%%%%%%%%%%%%%%%%%%%%%

\section{Introduction} 
\label{sec:intro}
X-ray binaries consist of a normal star and a collapsed compact object, either a black hole or a neutron star. These compact objects are fed by accreting matter from their donors due to  dominant gravitational pull from the compact star. 
%The in-falling matter forms an accretion disk as it spirals down to the compact object via losing its angular momentum. 
 The infalling gas heats up and emits radiation in the form of X-rays (see reviews by \citealt{remillard2006x} and \citealt{done2007modelling}). 
%vigorous collisions among the in-falling particles heat up the
%to $\sim10^{8}$K, and 
%This process eventually leads to conversion of gravitational energy into kinetic energy and subsequent 
%The properties of such systems have been reviewed in detail in

The X-ray binary systems are classified in two categories based on the type and mass of the companion star: High Mass X-ray Binaries (HMXBs) and Low Mass X-ray Binaries (LMXBs) \citep{lewin1997x}. The accretion in LMXBs generally proceeds via the Roche lobe overflow whereas for HMXBs, stellar winds from the massive companions play a vital role in accretion.  The HMXB's show the X-ray luminosity ($L_{x}$) of the order of $10^{35} -10^{40}$ erg s$^{-1}$ \citep{mineo2012x, ritter2003catalogue, liu2006catalogue}. 
%Such super-giant stars are known to show mass loss rate up to $10^{-6} M_{\odot}$ yr$^{-1}$
 Such super-giant stars are known to show mass loss rate between   $10^{-9}$- $10^{-5} M_{\odot}$ yr$^{-1}$ \citep{martins2015mass} through the stellar wind whose launching mechanism has been analyzed in details over the decades (\citealt{castor1975radiation,herrero1995fundamental, puls2006bright}; see also \citealt{2015A&A...575A...5C} for a recent review and 3-D modeling of the Cygnus X-1 wind). Only a fraction of the wind material reaches the compact object, and this fraction depends on the system geometry. Many of the systems are transient due to high ellipticity of the orbit.

Cygnus X-1 and HDE-226868 is an example of an HMXB system where HDE-226868 is a blue super-giant and the compact star is a black hole \citep{ziolkowski2014determination}. Observations show that the donor is orbiting the Cyg X-1 with a period of 5.6 days \citep{murdin1971optical, pooley1999orbital, ziolkowski2005evolutionary}.  The orbit is practically circular, so the source is persistent, but nevertheless it shows state transitions. The system is either in the Hard/Low state, in which the Comptonized power-law component  dominates, originating in a hot optically thin plasma at temperature of order of $10^8$ K, or in the Soft/High state, dominated by the cooler thermal disk component \citep{tananbaum1972observation, zhang19971996}. 
%The emission in the hard state of Cygnus X-1 is dominated by thermal Comptonization.
Although the exact geometry of the Comptonizing hot plasma in X-ray binaries is still unknown, most of the energy dissipation is expected to happen in the vicinity of the black hole \citep{fabian2015},  and the hard X-ray emitting site is usually refereed to as hot corona. This corona is sometimes described as a lamp-post model, particularly in the High/Soft state \citep[e.g.][]{fabian2015}, but, particularly in the Low/Hard state, it could be interpreted as the advection-dominated accretion flow (ADAF: \citealt{ichimaru1977bimodal, Narayan:1994xi, Rees:1982pe, Chen:1995uc}) or as a post-shock region in a centrifugally supported shock \citep{chakrabarti1997spectral}.

The transitions between the spectral states usually take a few days and the source remains in a specific state for a few months \cite{remillard2006x}. Sometimes the transition from the Hard to the Soft state is not fully performed, and the source remains in the intermediate state. Such transitions are referred to as "failed state transitions" \citep{pottschmidt2003long, grinberg2014long}. The observed X-ray variability properties of Cyg X-1 strongly depend on its spectral state (see e.g., \citealt{cui1997temporal} for a detailed study of the power density spectra corresponding to the state transitions).  In the Low/Hard state the broad-band power spectrum has a form of a few broad Lorentzian components.

Several studies have shown the presence of a focused wind by observing the source Cyg X-1 at various phase angles \citep{hanke2008chandra, Mi2016chandra}. This wind provides the material which accretes onto the central black hole.
The observed X-ray variability of Cygnus X-1 is most likely a combination of the propagating perturbations appearing at some characteristic radii of the hot flow. 
%focused by the accreting black hole's gravitational potential. 
In addition, there may be an important variability component due to the variable (clumpy) stellar wind from the companion \citep{fullerton2006discordance, Puls:2006fk}. These winds are powered by radiation pressure acting through absorption in spectral lines (CAK mechanism, \citealt{castor1975radiation}, see also \citealt{2015A&A...575A...5C} for a review.)
% The stellar wind from the supergiant provides the source of the matter accreting into the black hole and thus also affects the observed spectral states. The winds of luminous hot stars are driven by absorption in spectral lines and they are called line driven winds. This wind model has been extended by  Castor abott $\&$ Klein \citep{castor1975radiation} and now is commonly referred as CAK formalism employed to model the stellar wind in Cygnus X-1. Further investigations of such model are reviewed in \citep{2015A&A...575A...5C}.}
 %HDE 226868 shows a strong stellar wind. Such winds are driven by radiation pressure, due to copious absorption lines present in the ultraviolet part of the spectrum on material in the stellar atmosphere \citep{herrero1995fundamental}.} 
The structure of the stellar wind is an integral effect of the instability in the line driving force \citep{lucy1970mass, owocki1988time}. Simulations support the view that a steady solution of line-driven winds is not possible, i.e., perturbations are inherently present in the wind \citep{feldmeier1997possible}, causing variations in the parameters of the flow such as density, velocity and temperature, which compress the gas into small, cold, and overdense structures, often referred to as “clumps” \citep{oskinova2012clumped, sundqvist2013clumping}. Observational analysis also reveals and supports the view that this focused wind is quite clumpy depicting variations in density and velocity of the constituents of the flow. These clumps are stated responsible for the observed absorption dips (i.e the lower flux) in the Hard State spectra \citep{hirsch2019chandra, Mi2016chandra}. 
 %In addition to the complex wind profile, the clumpiness of the wind and its property are also subjected to a lot of discsussions\citep{fullerton2006discordance, Puls:2006fk}. 

%\textcolor{green}{\bf A recent study by \cite{2020arXiv200413241M} also discusses the soft and hard state spectrum from Cygnus X-1 and the clumpy stellar wind in the context of mass flow in the corona and accretion disk. In HMXBs, the wind dynamics such as wind terminal speed, are generally measured within  $\sim 20\% $ of the distance from the stellar surface.}

%{\bf The accretion rate tends to slow down in the inner hot region in X-ray binaries and are mostly studied as advection dominated accretion flows (Ichimaru 1977; Rees et al. 1982; Chenet al. 1995). Such flows have lower radiative efficiency than that of a standard, geometrically thin accretion disk due to advection of energy. The observed  spectrum  from the  accretion  disk is expected to have a thermal component and a power law tail representing the Soft and Hard state respectively. }

 In this work we aim to model in detail the short timescale variability ($<$100 s) of the flow in the Low/Hard state of Cygnus X-1, taking into account the time-dependent mass inflow from the companion star HDE-226868.
We represent the hot flow  performing the relativistic, hydro-dynamical, 2D model of accretion with the shock in a low angular momentum, transonic flow, described in detail in \cite{sukova2015oscillating,  sukova2017shocks}. 
We investigate the  oscillating shocks in the gas that passes through the multiple sonic points which may be the cause for enhanced emission of radiation from that region.  Our flow model is thus simpler than the two component accretion flow models \citep{Chakrabarti:1995mx, Chakrabarti:1997hs, cabanac2010variability} where a Keplerian disk with high viscosity at the equatorial plane is present inside a low angular momentum sub-Keplerian halo \citep{Kumar:2014iwa}.
Study of such models show that the observed variability of X-ray binaries are account of the oscillating hot coronal flow. 
However, the outer boundary is modeled by us in detail, to account for a varying density and angular momentum in the clumps, in accordance to the scenario of wind accretion in Cyg X-1.

\begin{figure}
\centering
\includegraphics[width = 1.0\textwidth]{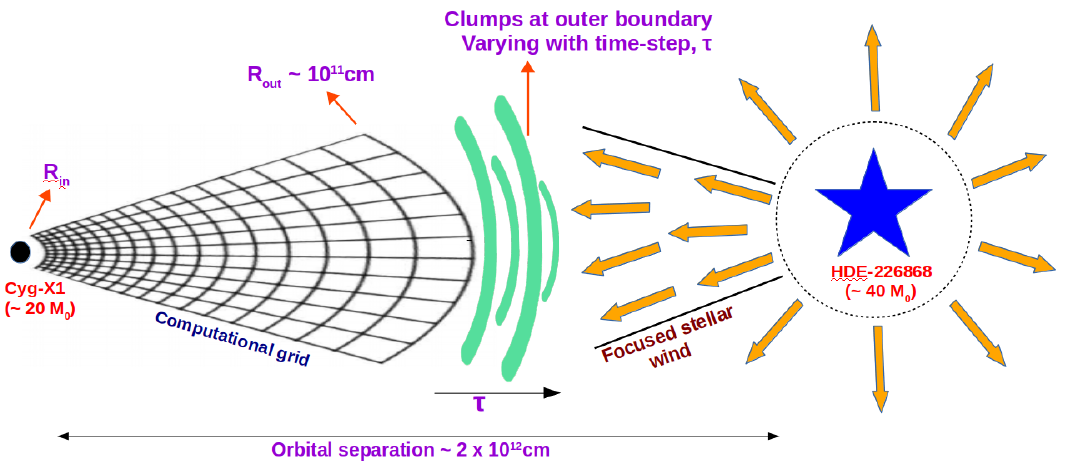}
\caption{The model set up for our simulation. The strips of clump are coming from the focused wind from the companion star at the outer boundary of our computational grid which reflects the outer part of accretion disk of the black hole in Cyg X-1. }
\label{fig:1}
\end{figure}

The article is organised as follows. Section \ref{Sec:2} provides the description of our model. The details of our initial set-up, time dependent outer boundary and the code used in numerical simulation are presented in \ref{Sec:2.1}, \ref{Sec:2.2} and \ref{Sec:2.3}, respectively. Our results are presented in Section \ref{Sec:3}. In \ref{Sec:3.1}, we show the comparative results between the model having time dependent and constant outer boundary conditions, and we discuss the nature and behaviour of the shock in our model. In \ref{Sec:3.2}, corresponding effects on mass accretion rate and power density spectra due to the shock oscillation are presented. In Section \ref{Sec:4}, we summarize and discuss our work and present the conclusions in Section \ref{Sec:5}.

%%%%%%%%%%%%%%%%%%%%%%%%%%%%%%%%%%%%%%%%%%%%%%%%%%%%%%%%%%%%%%%%%%%%%%%%%%%%%%%%%%%%%%%%%%%%%%%%%%%%

\section{Model Overview}
\label{Sec:2}
The Hard State of Cygnus X-1 is dominated by the hot flow, and the emitted radiation has the form of the hard X-ray power law component.  The observational constraints for the radial cold disk extension based on reflection and the geometry of the inner region are still under discussion (see \citealt{kara2019corona} and references therein). In our work, we focus on the dynamics of the hot flow and we neglect the interaction of the hot plasma with the cold disk.

We model an inviscid, non-magnetized, transonic accretion flow with a low angular momentum profile. The temperature in this  region tends to be very high due to a sudden decrease of the radial velocity and kinetic energy.  This model is very different from the high angular momentum single phase ADAF or the two-component cold disk plus high angular momentum coronal flow \citep[e.g.][]{rozanska2000,2020arXiv200413241M}, and rather follows the scenario proposed by \cite{chakrabarti1995spectral}.
Thus a hot, puffed up geometrically thick structure is formed where hot gas pressure and the centrifugal force together balance the gravitational force. Such centrifugally supported shocks caused by jump in the thermodynamic quantities have been already studied in detail by \cite{chakrabarti1995spectral} for optically thick and optically thin flows. They explain the power-law component observed in the spectrum due to Comptonization of cooler photons by hot electrons in the CENtrifugal barrier dominated BOundary Layer (CENBOL) which is a subsonic post-shock region close to the black hole. Further studies such as interaction of shocks with winds, spectral analysis of Cyg X-1 etc., are done based on the model considering Keplerian flow surrounded by sub-Keplerian halo \citep{Chakrabarti:1996ns, acharya2002interaction, mandal2007spectral}. Our model is to some extent similar to the CENBOL, but we don't model a thin disk. However, we consider the time-dependent evolution of the flow, with the effect of the wind clumpiness on the flow dynamics.

%Time dependent Angular momentum and Density at outer boundary
\begin{figure}
\includegraphics[width = 1.0\textwidth]{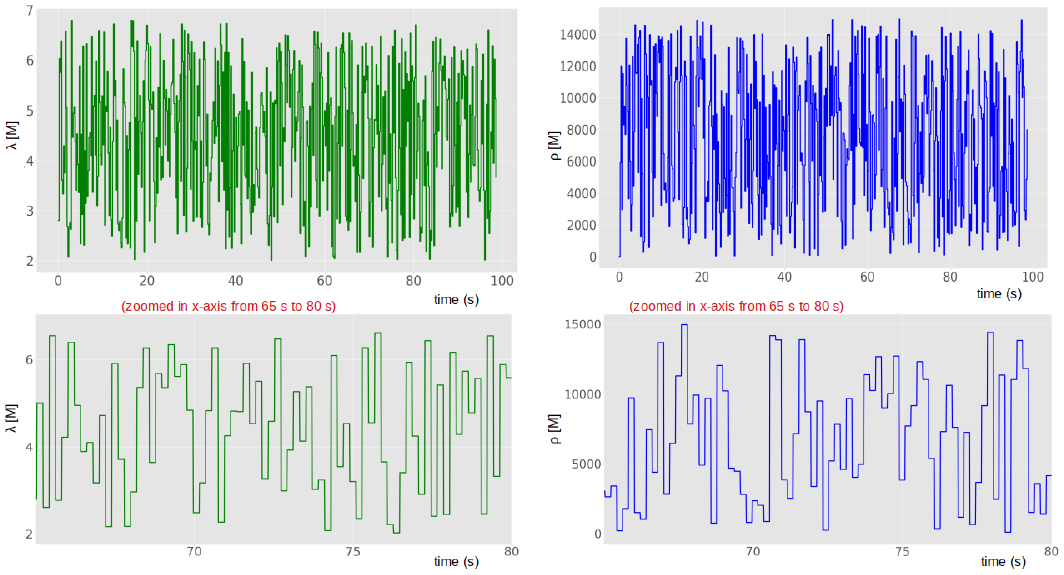}
\caption{Angular momentum (left) and density (right) at the outer boundary varying with time. The x-axis has been zoomed in to show the distribution more clearly. The minimum and maximum value for the distribution of angular momentum are 2.0[M] and 6.8[M] and that of density are 0[M] and 15000[M] respectively. }
\label{fig:2}
\end{figure} 

\subsection{Initial configuration}
\label{Sec:2.1}
The initial configuration roughly follows the concept of the quasi-spherical distribution of the gas, provided by constant specific angular momentum, $\lambda$, for a non-Keplerian accretion disk \citep{abramowicz1981rotation}. Such solutions are generally transonic, and both smooth transonic solutions and solutions with shocks exist, depending on the adopted parameters of the flow \citep[see e.g.][and references therein]{das2003,2012NewA...17..254D}. Here we focus on solutions with shocks. 

Our initial condition prescribes the shock solution for transonic accretion flow and the critical points of the flow are calculated at t = 0 s.  For such quasi-spherical flow, the mass accretion rate is derived from the continuity equation. Details are given in \citep{sukova2015oscillating,  sukova2017shocks, palit2019effects}. The choice of the parameters, ($\epsilon, \lambda, \gamma$) i.e., specific energy, specific angular momentum, and adiabatic index of the flow, determines the shock position and its nature.  We choose the parameter space in order to achieve an oscillating shock for a highly spinning black hole and not an accreting neither expanding shock. This enforces the variability in the mass accretion rate rather than a smooth inflow through the inner boundary. The equation for the critical points is based on \cite{das2002generalized} and the pseudo-Newtonian approximation for the Schwarzshild black hole. The location of shock in the Kerr metric can be very different from that in Schwarzschild regime. Thus we prescribe two values of angular momentum: $\lambda_{(\rm Schwarzschild)}$ to find the semi-analytic shock solution in non-spinning space time and $\lambda_{(\rm Kerr)}$ to prescribe rotation of the gas. The first one is used to determine the critical points and the Mach number profile at the initial time mainly to ensure that all three critical points exist, and the second one is the value, which is actually prescribed to the gas.

In our previous study \citep{palit2019effects}, we presented the models with low (in most cases, zero) spin only, so we assumed $\lambda_{(Schwarzschild)}$ = $\lambda_{(Kerr)}$. The shock existence for zero spin case (Schwarzschild), requires $\lambda \approx 3.3$[M] - 3.9[M] \citep{das2002generalized}. This value is used for the initial profile setting. The second value, $\lambda_{Kerr}$ has to be in the actual shock existence interval, which is much lower (and also very limited), e.g. around 2.8[M] \citep{sukova2015oscillating, 2012NewA...17..254D,  sukova2017shocks}. The evolution is governed by the actual gas angular momentum and settles into the appropriate solution.

Density ($\rho$), specific energy ($\epsilon$), radial velocity in Boyer Linquidst coordinate (U$^{r}_{BL}$), and entropy (K) are set at t = 0 s. We compute their profiles with $\epsilon = K \rho^{\gamma-1} / \gamma -1$ as the function of radius.  Here $K$ denotes entropy and is given by Equation 5 in \cite{palit2019effects}. The radial gradient for the flow velocity has been maintained finite in order to have a smooth and continuous accretion flow from the outer sonic point to the inner sonic point. We start our simulation with a quasi-spherical, non-magnetized, inviscid flow and we use an adiabatic equation of state with $\gamma= 4/3$. The density is calculated for such quasi-spherical flow from the continuity equation: $\dot{M} = \rho u r^{2}$ where $\rho$ is the density and u is the radial inward velocity. The choice of the value of $\dot{M}$ normalizes the density in the simulation and calculates the amount of density during the jump condition i.e at the shock at t = 0 s. Thus the amount of $\dot{M}$ we prescribe at initial time determines the amount of density in the post shock region. The value obtained for post shock density is higher for larger values of $\dot{M}$ given initially. Thus $\dot{M}$ is also our model parameter required in the calculation of the shock and sonic points at initial time. 

Figure \ref{fig:1} describes our set-up in the context of Cygnus X-1 binary system. In order to model such a system, we prescribe some parameters initially in our model from the estimations and observations available for this system. 
We choose the black hole spin $a=0.8$ \citep{kawano2017black} and outer radius of the computational grid, ($R_{\rm out}$) $\sim 10^{11}$cm. The outer boundary is chosen such that it is within Paczynski’s limit of the disk size suggested for Cyg X-1 \citep{paczynski1977model, romero2002recurrent}. Mass of the black hole Cyg X-1 and that of the super-giant \citep{mastroserio2019x, ziolkowski2014determination} are considered here in order to estimate the angular momentum and density at outer boundary. Their values are listed in Table \ref{tab:tab1}. In order to replicate the flow of clumps through a focused wind from companion star, we introduce a time dependent outer boundary condition where density and angular momentum are a function of time.
The resolution of our computational grid in the radial and azimuthal directions is 384 and 256 respectively.

 \begin{table}[ht]
    \begin{tabular}{|c|c|c|}
        \hline
       \large{Quantity}& \large{  Geometric unit [M]   }& \large{  Physical unit (cgs)} \\
       \hline
        \hline Mass of Cygnus X-1 & 1[M]  & 20 M$_{\odot}$  \\
        \hline Mass of HDE-226868 & -  & 40 M$_{\odot}$ \\
        \hline Radius of HDE-226868  & - & 20 R$_{\odot}$ \\
         \hline R$_{out}$ of computational grid & 50000[M] & $\sim 1.47\times10^{11}$ cm  \\
         \hline  R$_{in}$ of computational grid &  0.8[M] & {$ \sim 2.36\times10^{6}$ cm} \\
        \hline Shock location at t = 0 & 27.9[M] & $\sim 8.2\times10^{7}$ cm \\
         \hline Specific energy, $\epsilon$ & 0.005[M] &- \\
        \hline $\lambda_{Schwarzschild}$ & 3.4[M] &-\\
         \hline $\lambda_{Kerr}$ & 2.8[M]  &-\\       
         \hline $\lambda_{min}$   &  2.0[M] & -\\
         \hline $\lambda_{max}$  & 6.8[M] & -\\
         \hline $\rho_{min} $ & 0.0[M] & 0 g cm$^{-3}$ \\
         \hline $\rho_{max} $& 15000.0[M]& $\sim 10^{-15}$ g cm$^{-3}$ \\         
         \hline Time period, $\tau$ & 2000[M] & $\sim$0.1 s \\
         \hline Final time of simulation & $10^{6}$[M] & $\sim$100 s \\
         \hline Shock oscillation & 10.63[M] - 90.78[M] & 3.14$ \times 10^{7}$ cm - 2.68$\times 10^{8}$ cm\\
         \hline
    \end{tabular}
    \caption{Quantities used in our model, outcome from our simulation, and a few observed/estimated values of the source, Cygnus X-1. The values are given in geometric units, [M], and in the physical units. }
    \label{tab:tab1}
\end{table}
%%%%%%%%%%%%%%%%%%%%%%%%%%%%%%%%%%%%%%%%%%%%%%%%%%%%%%%%%%%%%%%%%%%%%%%%%%%%%%%%%%%%%%%%%%%%%%%%%%%%

\subsection{Time dependent outer boundary condition}
\label{Sec:2.2}

 Wind fed accretion models based on the Bondi-Hoyle-Lyttleton analysis has been studied in detail in the past (\cite{taam1989double}, see review by \cite{taam2000common}). With our time-dependent boundary condition, we study the relativistic, hydrodynamical hot inner flow in wind-fed geometry. We introduce clumps or strips of matter of varying density and angular momentum at the outer boundary, randomly changing with time. We chose a time period, $\tau$, which determines the interval of change of parameters at outer boundary. We selected the value of $\tau \sim 0.1$ s in such a way so that the update at outer boundary is neither too fast for the code to handle nor too slow to propagate the effect of the change. The final time of computation is $t_{\rm final}$ $\sim 100$ s. Such a choice of our time scales implies 1000 times update of the outer boundary during the evolution of flow.

 The shape of the clumps in our model is illustrated in Figure \ref{fig:1}. In our simulations the density and angular momentum at the outer boundary are only functions of time at the adopted outer radius, and not of the polar angle. Thus we ignore the fact that in  Cygnus X-1, the velocity field of the wind is very complex, and that the focused part of the wind accelerates in the direction towards the black hole. We also neglect the dependence of the measured radial wind velocities on the orbital phase  $\phi$ as studied in \cite{balucinska2000distribution}. The choice of elongated clumps maximizes the level of variability for a given cloud size in the radial direction. Larger but spherically symmetric clouds randomly distributed in the  $\theta$ direction would likely give the same effect. Hence our choice is convenient and maximizes the effect in an adequate amount of time evolution of the flow.

The prescription for density at the outer boundary is estimated from the observed data. We implement a function generating values between the prescribed maximum and minimum amplitude of density. The estimated values are at first converted from the physical units to dimensionless code units in order to prescribe in our simulation. The minimum value, $\rho_{\rm min}=0$ g cm$^{-3}$  i.e no inflow from the outer boundary and $\rho_{\rm max} = 10^{-15}$ g cm$^{-3}$ representing the maximum density of the clumpy matter present in the wind.
 A recent study by \cite{sundqvist20182d} reported the mass of clump as $10^{17}$ g and an average size of the clump as $\sim$ 0.01 $R_{\rm HDE}$ with density about $\sim 10^{-14}$ g cm$^{-3}$ ($R_{\rm HDE}\sim$ 20R$_{\odot}$ is the radius of the star, HDE-226868). We have smaller and less massive clumps in our model (size of our clumps $\sim$ 0.0001 $R_{\rm HDE}$) as compared to \citep{sundqvist20182d}. Our model lacks stratification in $\theta$ and $\phi$ directions, thus we choose smaller timescale of change at outer boundary to maximize the effect.
 Similarly for angular momentum prescription, we implement a function generating values between the maximum and minimum values. The choice is motivated by the values of angular momentum at marginally stable orbit, $\lambda_{\rm ms} =$ 3.67  $GM_{\rm BH} / c^{2}$ and at marginally bound orbit, $\lambda_{\rm mb} =$ 4  $GM_{\rm BH} / c^{2}$  \citep{das1999mass}. The minimum value $\lambda_{\rm min}$ is 2.0[M] and maximum value $\lambda_{\rm max}$ is 6.8[M], so that matter has angular momentum adequately higher or lower than the value at the marginally stable stable orbit, representing very fast and very slow rotation, respectively. We have tested our random number generator function and the distribution is uniform i.e the possible occurrence of any value within the given upper and lower bound is equally probable. We seed the generator to have the same starting point so that we get the same sequence of numbers for each run. Figure \ref{fig:2} shows the generated values of angular momentum and density at outer boundary.  The inflow of clumps from the companion star in Cygnus X-1 system, has not been identified having any particular pattern so our choice of such random distribution is quite reasonable. 

%%%%%%%%%%%%%%%%%%%%%%%%%%%%%%%%%%%%%%%%%%%%%%%%%%%%%%%%%%%%%%%%%%%%%%%%%%%%%%%%%%%%%%%%%%%%%%%%%%%%

%\subsection{Numerical Modelling}
\subsection{Time-dependent evolution}
\label{Sec:2.3}
We are using the numerical code HARM which is a conservative, shock-capturing scheme to solve numerically the continuity equation, the four-momentum-energy conservation equation, and induction equation in the GR framework \citep{gammie2003harm}. The numerical code follows the dynamical evolution of the flow by solving numerically the continuity Eq.(2), the four-momentum-energy conservation Eq.(4), and induction Eq.(18) mentioned in \citep{gammie2003harm}, in general relativistic framework. In our model, we are dealing with non-magnetized flow so we switch off the prescription of magnetic field in the code and thus integrated form of equation only considers the continuity and momentum- energy conservation equation. HARM solves GRMHD equations in the modified version of the Kerr-Schild coordinate system (KS) rather than Boyer-Lindquist coordinates thus matter can accrete smoothly through the horizon and evolution of the flow can be followed properly without reaching any coordinate singularity. Having a logarithmic mapping of the radial component of the grid helps to zoom in close to the black hole. 2D simulations are axisymmetric, i.e. the derivatives of quantities in $\phi$ direction are neglected (although, velocity field and magnetic field vectors still have all the 3 components). The publicly available version of the code has a free inflow-outflow boundary condition \citep{JSFI177}. The constant mass inflow rate at the outer boundary condition was used earlier in \citep{palit2019effects}. In the present simulation, the code has been further supplied with the time dependent outer boundary condition.

%%%%%%%%%%%%%%%%%%%%%%%%%%%%%%%%%%%%%%%%%%%%%%%%%%%%%%%%%%%%%%%%%%%%%%%%%%%%%%%%%%%%%%%%%%%%%%%%%%%%

\section{Results}
\label{Sec:3}

\begin{figure}[ht]
\centering
\includegraphics[width = 1.0\textwidth]{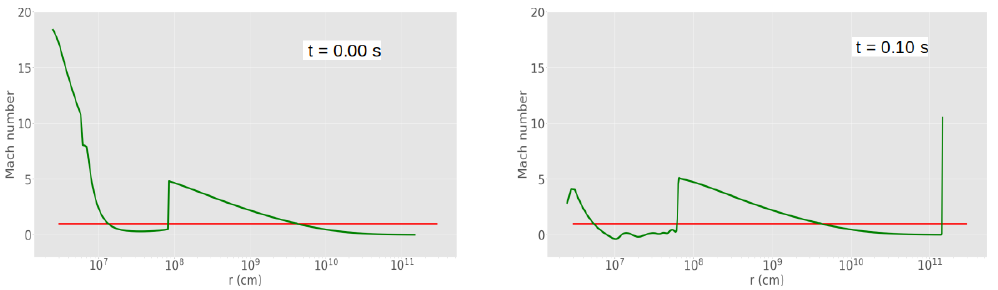}\\
\includegraphics[width = 1.0\textwidth]{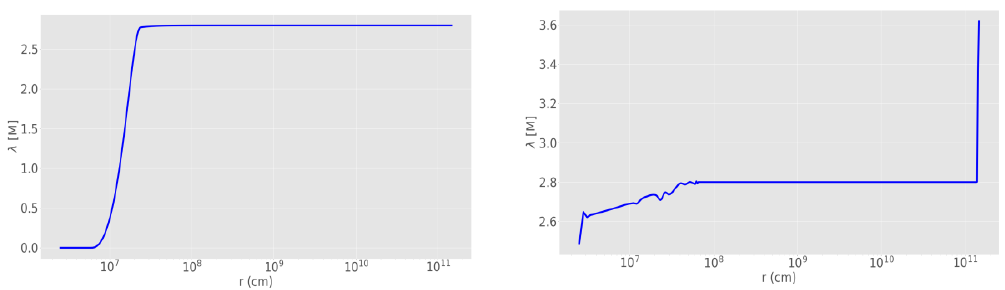}\\
\includegraphics[width = 1.0\textwidth]{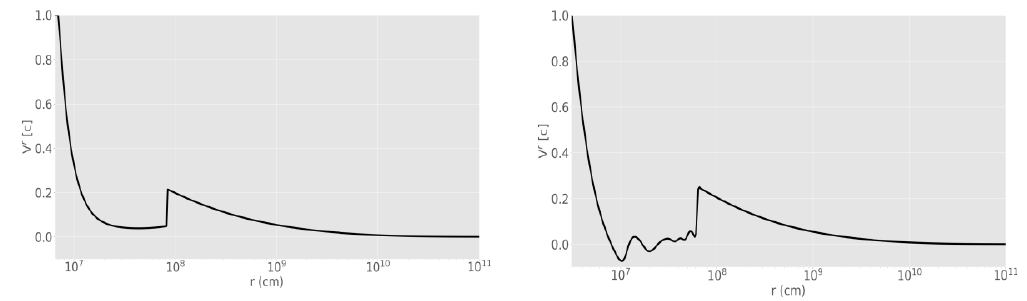}\\
\includegraphics[width = 1.0\textwidth]{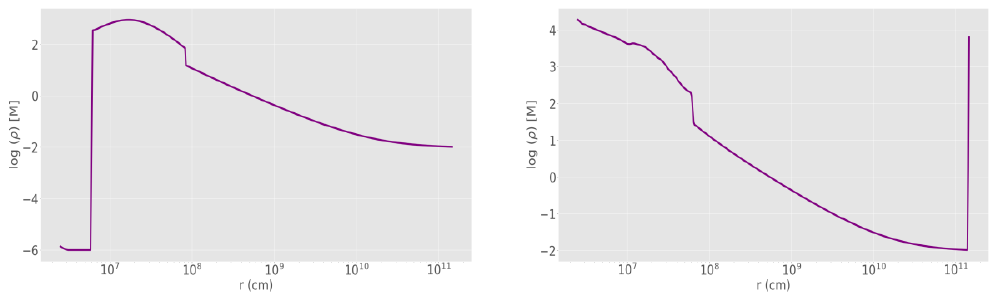}
\caption{Equatorial slice of Mach number, specific angular momentum, radial velocity and density at t = 0.0 s and t = 0.1 s showing update at outer boundary. The time period of boundary update is 0.10 s.}
\label{fig:3}
\end{figure} 

\begin{figure}[ht]
\includegraphics[width = 1.0\textwidth]{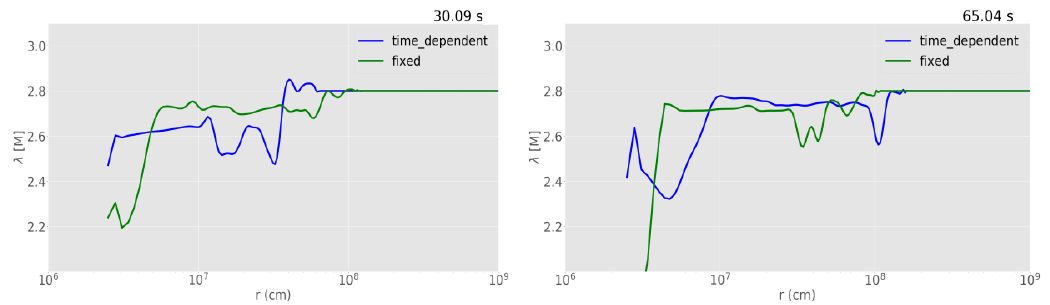}
\caption{Equatorial slice of angular momentum at t = 30.09 s and t = 65.04 s. The green line is from the simulation with fixed outer boundary conditions and the blue one is from the time dependent outer boundary update. }
\label{fig:4}
\end{figure}

\begin{figure}
\includegraphics[width = 1.0\textwidth]{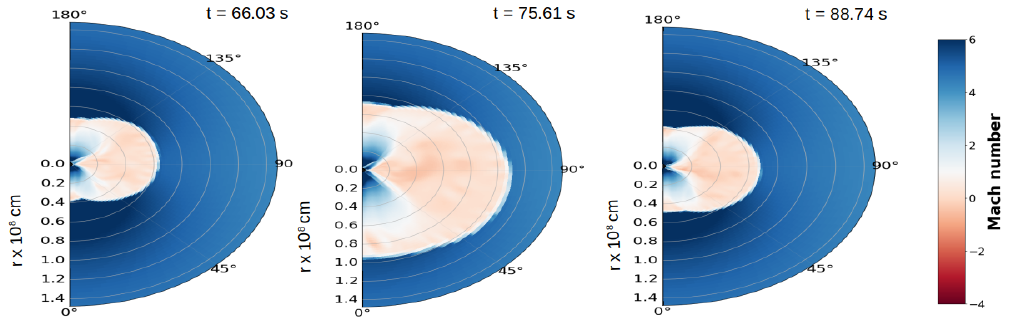}\\
\includegraphics[width = 1.0\textwidth]{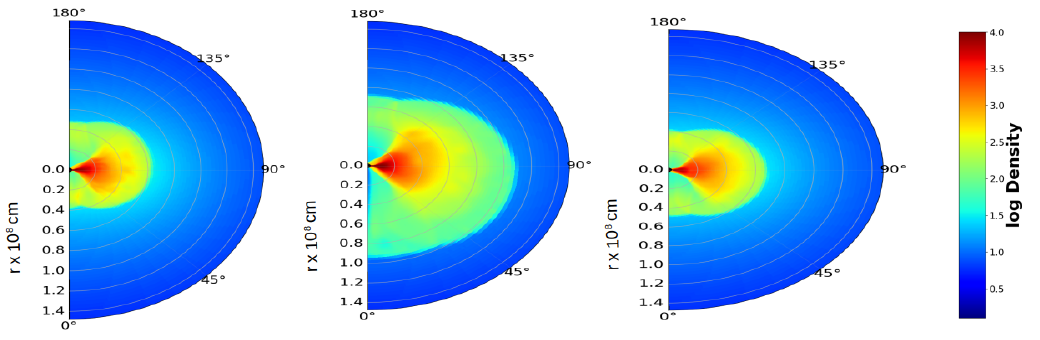}\\
\includegraphics[width = 1.0\textwidth]{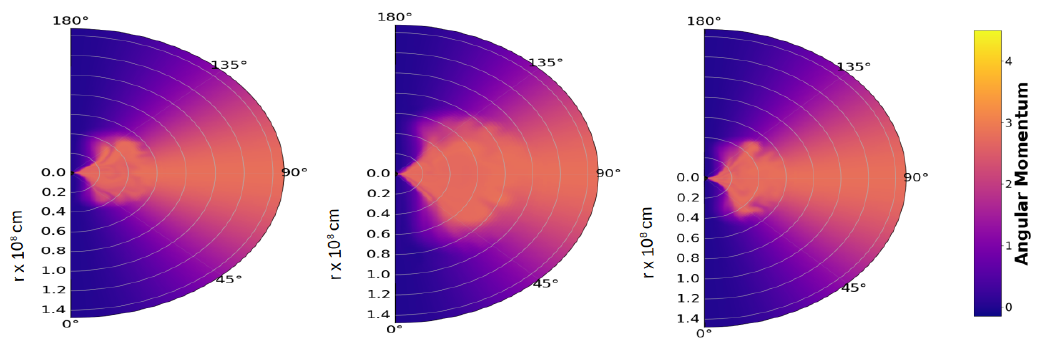}
\caption{Distribution of the Mach number, density and specific angular momentum, taken at t = 66.03 s, 75.61 s and 88.74 s, from left to right. The change of the size of the shock bubble can be seen in the different time snapshots. }
\label{fig:5}
\end{figure} 

Below we present the results of our simulations. The computations are done using dimensionless units, and later converted to the physical units. Table \ref{tab:tab1} lists the values of the parameters used in the model and some as the output from the computation. The radial and time scale in all the plots are presented in cgs units\footnote{The units in the numerical simulation are geometrical (G = c = 1, [r] = [t] = $[\lambda]$ =[M]). Time unit conversion: t=1[M] equals to time needed by light to travel half of the Schwarzschild radius of a black hole, hence the conversion factor: GM$_{\rm Cyg}/c^{3}$. Similarly, the conversion factor for radius from the code units in [M] to standard physical units is GM$_{\rm Cyg}/c^{2}$ (gravitational radius).}. Results from our simulation with modified boundary conditions show that all the physical quantities of the flow such as density, angular momentum, radial velocity, and Mach number, are modified quantitatively compared to the results obtained with a fixed outer boundary condition.

%%%%%%%%%%%%%%%%%%%%%%%%%%%%%%%%%%%%%%%%%%%%%%%%%%%%%%%%%%%%%%%%%%%%%%%%%%%%%%%%%%%%%%%%%%%%%%%%%%%%

\subsection{Shock behaviour and effects due to changes at the outer boundary}
\label{Sec:3.1}
Below, we present the 1D (equatorial) and 2D profiles of the Mach number, density, angular momentum and radial velocity. The effects of the update at the  outer boundary is seen after 0.01 s, when the perturbation starts propagating. At time t=0, the shock is located at $\sim$ 8.2$\times10^{7}$ cm. Figure \ref{fig:3} shows the jump at the outer boundary in the Mach number, angular momentum, radial velocity and density, taken in the equatorial plane, at t=0.1 s. During the dynamical evolution, the angular momentum of the gas close to the horizon should drop as it has passed through the inner sonic point. This can be seen in the equatorial snapshots of angular momentum. The red line in Figure \ref{fig:3} for the Mach number plots represents Mach number = 1, i.e the flow radial velocity is equal to the local sound speed. Region above this red line represents supersonic flow and below this represents the subsonic flow.

Figure \ref{fig:4} shows the angular momentum profile at the equator, for two cases of outer boundary conditions: fixed and time-dependent. The X-axis is zoomed in this figure in comparison with Figure \ref{fig:3}. The choice for snapshots at t = 30.09 s and t = 65.04 s is motivated by the peaks in the mass accretion rate which are seen at such times (see section \ref{Sec:3.2}). The value of $\lambda$ at the outer boundary at t = 30.09 s is 4.5[M] and at t = 65.04 s is 4.0[M] for the time dependent outer boundary case, whereas for the fixed outer boundary case, the value of $\lambda$ is constant, 2.8[M] all the time during the evolution.  The $\lambda$ around the inner sonic point at t = 30.09 s drops to 2.4[M] (time-dependent outer boundary) and 2.1[M] (fixed outer boundary) and at t = 65.04 s, $\lambda$ is  2.3[M] (time-dependent outer boundary) and 1.7[M] (fixed outer boundary). Also the shock position is shifted due to the time dependent outer boundary condition. 

\begin{figure}
\centering
\includegraphics[width = 0.8\textwidth]{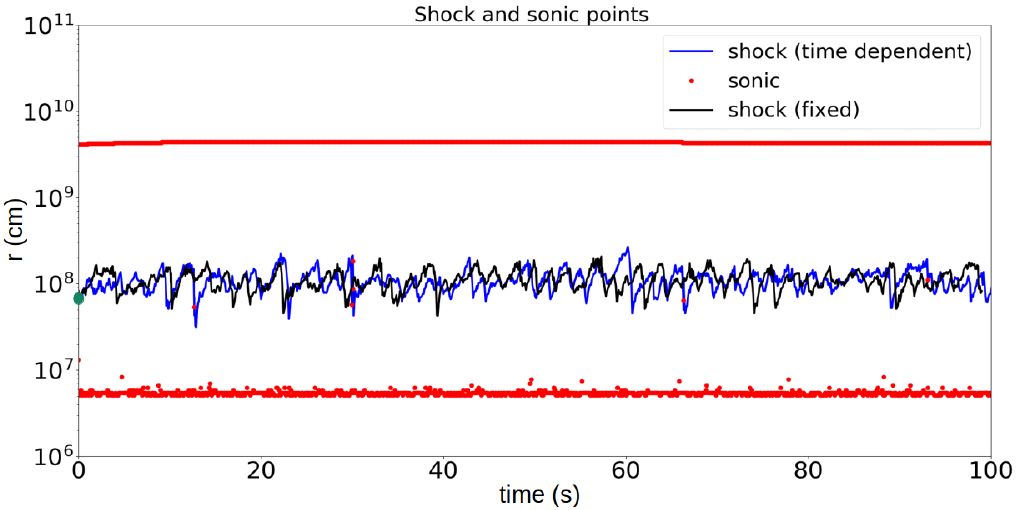}
\caption{Position of shock, inner sonic and outer sonic points varying with time for the time dependent and fixed outer boundary conditions. The location of shock at t = 0 s is marked with green dot on y-axis.}
\label{fig:6}
\end{figure}
The shock formed in our model is oscillatory, as expected \citep{sukova2017shocks}. The shock is oscillating between r $\sim 3.14\times10^{7}$ cm and r $\sim 2.68\times10^{8}$ cm. The plot shows regions representing supersonic (blue) and subsonic flows (red) as seen in the 2D Mach number distribution. Figure \ref{fig:5} shows the changing size of the shock bubble in different time snapshots, as seen within the Mach number and density profiles. The angular momentum plots show the gas with low value of angular momentum, accumulated around the poles, and the the maximum amount in the subsonic region. The subsonic region appears to be more dense compared to the supersonic regime thus slowing down the rotation of matter in the shock bubble. 

Figure \ref{fig:6} shows the oscillating shock during the evolution of the flow till the final time of our simulation i.e 100 s. We compare here two cases, the time dependent and the fixed outer boundary conditions, using blue and black lines, respectively. The location of the shock position varying over time for both cases are not exactly same and can be clearly seen in the figure.

\begin{figure}
\includegraphics[width = 1.0\textwidth]{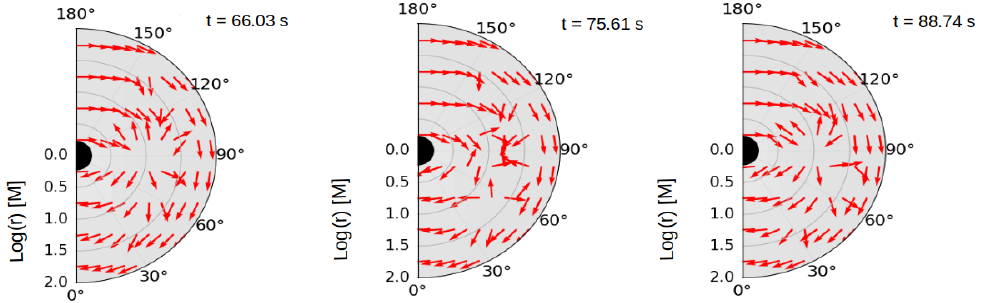}
\caption{{ Velocity field with zoomed logarithmic radial axis. The time snapshots are the same as in Fig. \ref{fig:5}, to show the behaviour during the oscillation of the shock bubble.} }
\label{fig:7}
\end{figure}

  Figure \ref{fig:7} shows the velocity field indicating the direction of the particles around the shock. (Note that the vectors of velocity field are normalized to unity.) The radial scale is in geometric units and is logarithmic in this figure. The oscillation of the shock is between r = 10.63[M] (3.14$ \times 10^{7}$ cm) and r = 90.78[M] (2.68$\times 10^{8}$ cm). In geometric units, it corresponds to 1.0[M] and 1.9[M] on log scale thus the radial scale is 0.0[M] - 2.0[M] in figure \ref{fig:7}. The time snapshots in Figure \ref{fig:7} are same as in Figure \ref{fig:5}. It can be seen that at t = 66.03 s and t = 88.74 s, the gas moves outwards from the shock, whereas at t = 75.61 s the velocities are pointing inwards. By comparison with the density distributions plotted in Figure \ref{fig:5}, we show that when the density is low, 
%the gas moves outwards from the shock, indicating that 
the shock is pushed away.
%from being accreted. 
On the other hand, when the density is high, the velocity of the flow changes direction and the gas moves inward, implying that the pressure waves are pushing the shock inward. These few snapshots illustrate how the shock bubble moves back and forth throughout the simulation.

\begin{figure}
\includegraphics[width = 0.7\textwidth]{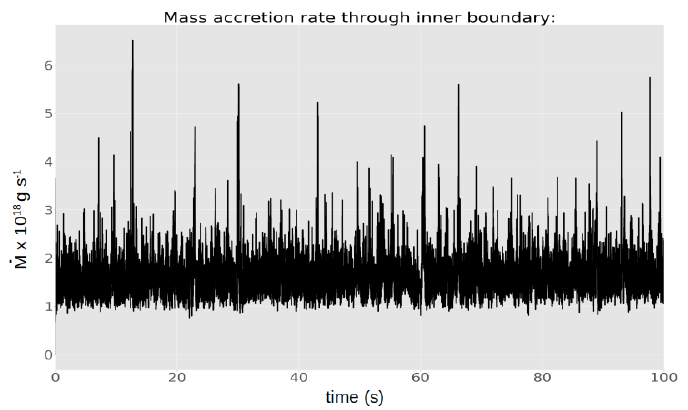}\\
\includegraphics[width = 1.0\textwidth]{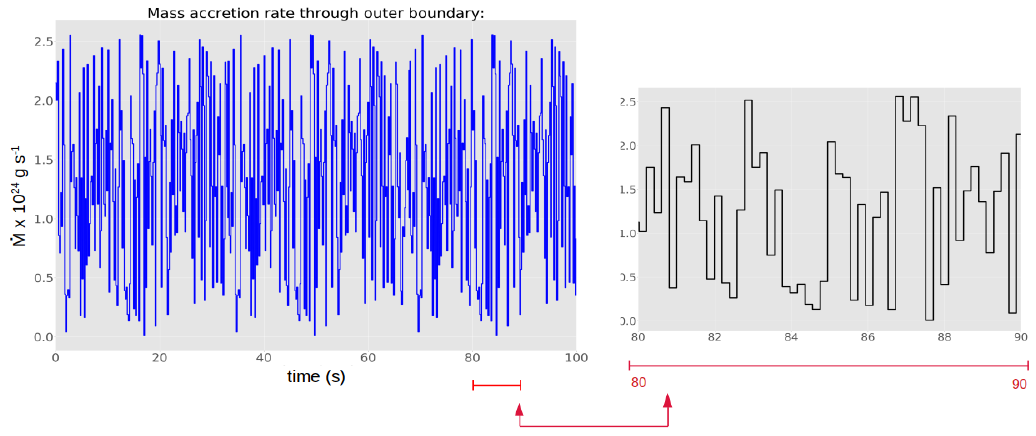}
\caption{The mass accretion rate calculated through the inner and outer boundaries. The x-axis and y-axis have been converted to physical units for direct comparison to observed quantities. The mass accretion rate through outer boundary has been zoomed to
$\Delta t= 80-90$ s in a small insert.}
\label{fig:8}
\end{figure} 

%%%%%%%%%%%%%%%%%%%%%%%%%%%%%%%%%%%%%%%%%%%%%%%%%%%%%%%%%%%%%%%%%%%%%%%%%%%%%%%%%%%%%%%%%%%%%%%%%%%%%

\subsection{Mass accretion rate and Power Density Spectral (PDS) analysis}
\label{Sec:3.2}
 We have tested that if the shock gets accreted through the inner sonic point, there is no variability in the mass accretion rate and it becomes flat throughout the evolution. So  it is important to have a low value of the specific angular momentum to avoid the shock from being accreted and to see the variability in the light curve due to the oscillation of the shock.
Figure \ref{fig:8} shows the mass accretion rate calculated in the inner part at 3$r_{g}$ and at outer boundary of our computational grid in the first and second frame, respectively. In the second frame, the outer boundary mass accretion rate has been zoomed in the inset to show the effect of our boundary update (see second frame of Figure \ref{fig:2} above, showing the density at outer boundary). The variability in the inner mass accretion rate can be seen here with many high peaks, as expected for accretion in the X-ray binaries mostly fed by focused winds. The observed mass accretion rate of our considered source, Cygnus X-1, is $\sim$ 0.02$\dot{M}_{Edd}$ $\sim$ 2.57$\times10^{17}$ g s$^{-1}$ and is quite close to the accretion rate that we have in the model. 

\begin{figure}[ht]
\includegraphics[width = 0.7\textwidth]{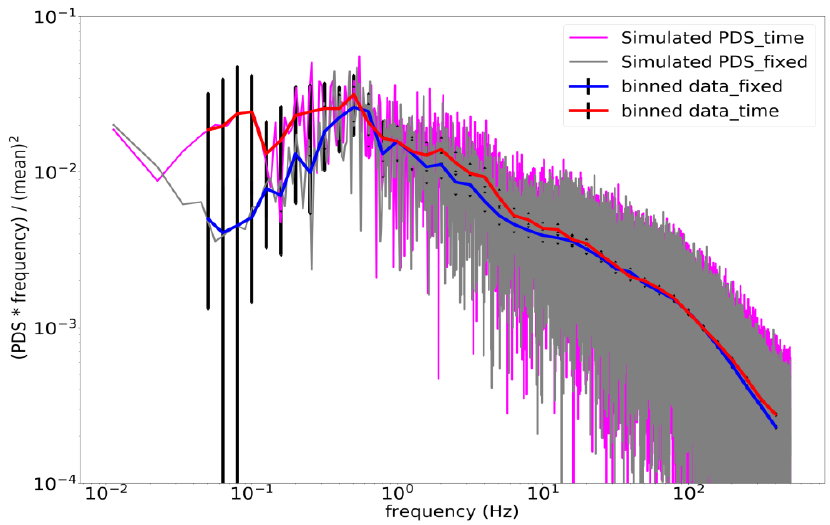}
\caption{ Power Density Spectra (PDS) calculated from simulated light curve for both the models: fixed and time dependent outer boundary with their respective logarithmically binned data.}  
\label{fig:9}
\end{figure}

 Table \ref{tab:tab2} contains the comparison between different parameters for the two types of boundary condition.
Figure \ref{fig:9} shows the power density spectral analysis (PDS) for both cases and the Lorentzian fitting results for the model with time dependent outer boundary.  The amplitude of the spectrum and the fractional variability of the light curve in the time-dependent model differs from that obtained with the fixed outer boundary condition and the values are given in Table \ref{tab:tab2}. The fractional variability is calculated as: 
\begin{equation}
   F_{var} = \sqrt{Variance} = \sqrt{ \frac{1}{n} \sum_{0}^{n} \frac{(x - x_{avg})^{2}}{x_{avg}^2} }
   % \label{eq:1}
\end{equation}
{ where x is the obtained light curve in our simulation.}
%The PDS we obtain from our simulated data has a lot of noise at high frequencies. To filter out our PDS in the latter half of the plot,
We imposed the logarithmic binning of our data and plotted the values averaged over 40 bins. The error bars at low frequencies are large because the data are largely spread around the mean value, but gradually the error bars decrease, giving perfect binned data. The calculated power density spectrum is further re-normalized to units of the squared mean of the light curve and then multiplied by frequency. This helps in better visualization of the distribution of power over the Fourier frequencies. Following \cite{pottschmidt2003long}, we fitted the PDS shape with the Lorentzian function, using their formula:

%\textcolor{red}{Y-axis in Figure \ref{fig:9} and \ref{fig:10} }are also divided by the square of mean error so that we have proper normalization. 
\begin{equation}
    L(f) = \pi^{-1} \frac{2R^{2}Q f_{r}}{f_{r}^{2} + 4Q^{2}(f-f_{r})^{2} }
    \label{eq:1}
\end{equation}
where R is the normalization constant, Q is the width of the Lorenztian and $f_{r}$ is the resonance frequency.

\begin{figure}[ht]
\includegraphics[width = 1.0\textwidth]{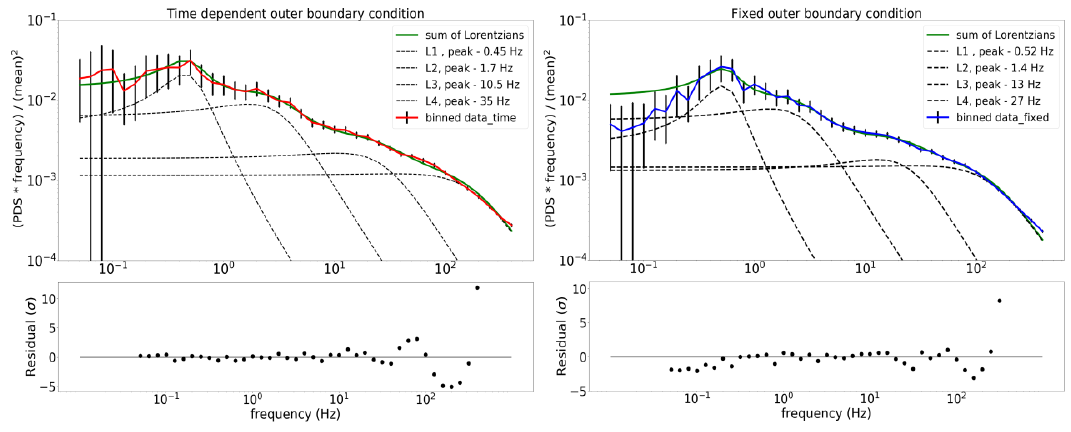}
\caption{ Four Lorentzian fit in the binned PDS for the time dependent outer boundary case and fixed outer boundary case in right and left panel respectively.} 
\label{fig:10}
\end{figure}

We use four broad Lorentzian curves, which is enough to get a good fit.  Figure \ref{fig:10} shows the Lorentzian fitting on the binned PDS for both the time dependent and fixed boundary condition models. The residual panels are also added in figure \ref{fig:10} for both the models. In our PDS, Lorentzians are labelled as L1, L2, L3, and L4. We allowed the free parameters to vary, i.e. normalization constant, the width of Lorentzian function, and the apparent peak frequency from the PDS for each of these four profiles, and we tested a set of 5 suitable values for each parameter. We calculated the minimum of chi-square distribution. 

The best fit $\chi^{2}$ value and the reduced chi-squared value ($\chi^{2}$/degrees of freedom) for the model with time dependent outer boundary conditions is  247 and 247/(40-12) = 8.84 respectively. It can be seen that the time dependent model provides a better fit in the low frequency part than the fixed boundary case. The $\chi^{2}$ values are high in our models due to very small statistical errors in the high frequency part in the PDS. They are actually dominated by the highest frequency point. Below 100 Hz the deviations in individual bins are of order of 2 $\sigma$ or smaller, very similar to the data residuals shown in Fig.~12 of \citet{pottschmidt2003long}.
%we have 40 number of bins for both the time dependent and fixed boundary models.
It can be inferred from the Lorentzian fitting in Figure \ref{fig:10} and Table \ref{tab:tab2} that the lower values of Q result in broader Lorentzians and vice versa. The broader Lorentzians imply less resolved part of the disc. The corresponding peak frequencies for time varying outer boundary are 0.45 Hz, 1.7 Hz, 10.5 Hz, and 35.0 Hz. They are very close to the frequencies obtained in \cite{pottschmidt2003long} for Cygnus X-1 hard State i.e. 0.2Hz, 2Hz, 6Hz and 40 Hz, and also close to the values of 0.3 Hz, 2.3 Hz and 9.3 Hz, reported by \citet{axelsson2018breaking} in their three Lorentzian fit (L1,L2, L3). Also our modeled variability level matches quite nicely the observations of Cygnus X-1 in its Hard State, presented in Figure (2.a–c) in \cite{pottschmidt2003long}.  The normalization constant, R is correlated with the peak frequency, $f_{r}$ in Eq. (\ref{eq:1}). This agrees with the observed PDS and R$ - f_{r}$ relation shown in \cite{pottschmidt2003long} for the Hard State.  The Lorentzian fitting shows that the hot flow close to the black hole is mostly similar for both models but the variability increases a little in the case of time varying outer boundary case. The major difference can be seen in the first Lorentzian which represents the oscillations at large radii. The amplitude of the Lorentzian is higher for time dependent case showing higher variability at this region compared to the fixed outer boundary condition. This is what we expect due to the changes at he  outer boundary.

 \begin{table}[ht]
    \begin{tabular}{|c|c|c|}
        \hline
       \large{Parameters }& \large{ Fixed outer boundary  }& \large{Time dependent outer boundary} \\
       \hline
         \hline $\dot{M}_{mean}$&   37109.041[M]&      37116.388[M] \\
         \hline $\Delta$$r_{shock}$& 55.828[M]& 80.1502[M]\\
         \hline ($\dot{M}_{max}$ -$\dot{M}_{min}$ )/ $\dot{M}_{mean}$& 3.35& 6.54\\
         \hline $F_{var}$&   0.216& 0.251 \\
         \hline Peak frequency (L1)&  0.52 Hz& 0.45 Hz \\
         \hline Peak frequency (L2)&   1.4 Hz& 1.7 Hz \\
         \hline Peak frequency (L3)&  13 Hz&  10.5 Hz \\
         \hline Peak frequency (L4)&   27 Hz& 35 Hz\\
         \hline Amplitude (L1)&  3.9e3& 5.04e3\\
         \hline Amplitude (L2)& 1.6e4& 1.8e4\\
         \hline Amplitude (L3)& 2.35e4&  3.5e4\\
         \hline Amplitude (L4)& 9.3e4&  9.5e4\\
        \hline Width (L1)&  1.1& 0.8\\
        \hline Width (L2)&  0.09& 0.1\\
        \hline Width (L3)& 0.09& 0.04\\
        \hline Width (L4)& 0.01& 0.01\\
        \hline  $\chi^{2}$& 247.60 & 427.03\\
        \hline  Reduced  $\chi^{2}$&  8.84 &   15.25\\
         \hline
    \end{tabular}
    \caption{Table contains the mean mass accretion rate, range of oscillation in radius, and the parameters of the 4 lorentzians for both types of boundary condition. Also the values of fractional varibility and chi-squared values are listed.}
        \label{tab:tab2}
\end{table}
%%%%%%%%%%%%%%%%%%%%%%%%%%%%%%%%%%%%%%%%%%%%%%%%%%%%%%%%%%%%%%%%%%%%%%%%%%%%%%%%%%%%%%%%%%%%%%%%%%%%%

\section{Discussion}
\label{Sec:4}
In this work, we model the focused wind in the binary system consisting of Cygnus X-1 and HDE-226868. We compute the structure and evolution of the hot accretion flow in the vicinity of the black hole. This plasma is responsible for the emission of hard X-rays which dominates the spectral appearance of the source in the Hard State. The wind inflow forming our outer boundary conditions is prescribed by the random changes of the angular momentum and density that are captured by the accretion flow from the companion. We follow the evolution of mass accretion rate in such flow and we calculate the power density spectra from our simulated light curve. 

The modeled PDS is well fitted by four Lorentzian curves. The values of the peak frequencies in our model and the overall normalization of the derived power spectrum matches well the observed power spectra of Cygnus X-1 in the Hard State \citep{pottschmidt2003long,axelsson2018breaking}. Note that \citet{axelsson2018breaking} attribute the peaks of the Lorentzians to the specific regions in the inner hot accretion flow, as illustrated in their Figure 5. They draw their conclusion on the basis of frequency-resolved spectra which allow to see the connection between the spectrum hardness and a specific Lorentzian component.  In our study, we don't have any thin disc like in \citet{axelsson2018breaking} but only hot accretion flow. The light curve, and thus the our PDS, come directly from the accretion rate calculated at inner part of the flow. The inner flow carries the history of the perturbation both supplied at outer boundary, and spontaneously created within the flow, mostly close to the shock position.

Our model cannot directly address the issue of the location of the specific emission component since the emissivity properties of the hot medium are not yet included in our model. On the other hand, we follow numerically the flow dynamics so all the aspects of the variability model based on the idea of the propagating fluctuations \citep{lyubarskii1997,kotov2001,ingram2012} are automatically included by us. We do not have to restore to semi-analytical approach as done for example by \cite{ingram2013exact} who use analytic expressions for the power spectrum for models with propagating fluctuations but with zero-centred Lorentzian of a width
$1/t_{visc}$, or by \cite{ahmad2018modeling} for the geometrically thin disc who modelled the variability characteristic in the Soft State.

The model we use is appropriate for black hole accretion due to the specific inner boundary conditions. As shown by \cite{2000A&A...358..617S}, the behaviour of power spectrum differs between neutron star and black hole systems. The power density spectra for black hole binaries show a rapid decrease at frequencies above 10 Hz in the low spectral state. 
Enhanced variability of the sources with neutron stars at the shortest frequencies are likely related to the presence of the boundary layer between the accretion flow and a star in these sources \citep[e.g.][]{popham2001}.

Our time-dependent evolution of the hot flow does not yet include the interaction with the cold disk which likely overlaps with the hot flow at some range of radii even in the Hard State \citep[e.g.][]{basak2017analysis, zdziarski2020two}. This interaction can lead to cooling of the hot flow, to the coronal flow - disk flow mass exchange as well as the transfer of the angular momentum from the hot flow to the disk. In addition, X-ray irradiation from the hot coronal flow can illuminate the outer disc and is able to drive thermal winds as outflows which are also able to extract angular momentum \citep{dubus2019impact}. Also magnetized winds are capable of such angular momentum transport. Out of these all effects, the issue of the angular momentum budget is the most important one for our model since the existence of the shock requires the angular momentum removal to remain in the proper parameter space. However, the dynamics of the cold flow and, better, the strongly stratified flow is far more difficult to model since it requires advanced treatment of the cooling, high spatial resolution and very time consuming computations due to the very high dynamical range of the involved timescales. Very preliminary results for such global simulations were achieved so far by \citet{Jiang2019}, for two sets of parameters that are appropriate for active galactic nuclei.  

 Our model is a very simple attempt to study the clumpy flow into Cygnus X-1. Here, our clumps are not $\theta$ dependent and are smaller in size than the estimated size of clumps via observational analysis \citep{sundqvist20182d}. This results in very small effect of perturbation at the outer boundary, propagating into the inner part. It is possible to have bigger clump size with larger time period of change but that would require more complex numerical analysis.  
We plan to implement the $\theta$ dependence of outer boundary inflow in the future work. On the other hand, it is important to note that the normalization of the high frequency tail of our normalized PDS is in rough agreement with the observational data. This may indicate that these high frequencies are indeed related to the dynamical instabilities close to the shock location while the wind perturbations are responsible for the longest timescales. The Lorentzian at $\sim $0.5 Hz is affected most strongly by the wind-related variability, both in position and - even more - the width. This Lorentzian gets even broader but remains at the same position when the source is on its way to changing the state to the High/Soft one \citep[see e.g.][]{pottschmidt2003long}. This is consistent with the universal wind-based mechanism at the longest timescales.

%\begin{figure}[ht]
%\includegraphics[width = 0.5\textwidth]{./figures_new/disc_1.png}
%\caption{The iilustration to show different peak frequency corresponds to oscillations in different part of the hot accretion flow onto the black hole. }
%\label{fig:10}
%\end{figure} 
%%%%%%%%%%%%%%%%%%%%%%%%%%%%%%%%%%%%%%%%%%%%%%%%%%%%%%%%%%%%%%%%%%%%%%%%%%%%%%%%%%%%%%%%%%%%%%%%%%%%%

\section{Conclusion}
\label{Sec:5}
 We conclude that our low-angular momentum coronal accretion model is able to represent the propagation of the perturbation from the outer boundary and explain the variability pattern seen in the Hard State of Cygnus X-1.
 %further affects the shock oscillation and angular momentum profile. 
 Power density spectrum (PDS) analysis of our simulated light curve shows close resemblance to the observed Hard State PDS of this source.  We model the PDS quantitatively by Lorentzian fitting and calculating the peak frequency of oscillations. Both the overall normalization of the PDS and the frequencies of the Lorentzians agree with the data analyzed by \citet{pottschmidt2003long}. The variability at peak frequencies corresponds to different parts of the disk.  The time-dependent boundary condition due to the wind clumpiness affects predominantly the lowest frequencies, below 1 Hz, while the higher frequencies are originating closer to the black hole in the flow itself. Their appearance is related to the shock formation in the low angular momentum flow. Assuming much lower angular momentum corresponding to the solution without a shock leads to a smooth inflow with no variability effects. Since the focused clumpy wind accretion characteristic for Cygnus X-1 is reflected in the PDS only at the lowest frequencies, the hot flow model discussed in this paper may also apply to LMXB which show basically similar PDS \citep{2000A&A...358..617S}. Future research should address the issue of the hot flow/cold disk interaction and the efficient removal of the angular momentum from the hot flow.
 %Comparing our results to the observed data we conclude that our frequencies correspond to hot inner accretion flow close to the black hole where the powerful X-ray emission arises.  {\bf The observed Power spectra for low mass X-ray binaries  are also possible to be analyzed by  If such hot flow develops in the case of LMXBs too, it is possible to explain the power spectra in such case with our fixed outer boundary model. These oscillatory shocks which are characteristic to hot flow might not effect the x-ray emissivity but it will affect the propagation and accretion flow in LMXBs.}

 %Perturbation from star makes the normalisation higher.
%%%%%%%%%%%%%%%%%%%%%%%%%%%%%%%%%%%%%%%%%%%%%%%%%%%%%%%%%%%%%%%%%%%%%%%%%%%%%%%%%%%%%%%%%%%%%%%%%%%%%

%\subsection{Discussion}
%\label{Sec:4.3}

%%%%%%%%%%%%%%%%%%%%%%%%%%%%%%%%%%%%%%%%%%%%%%%%%%%%%%%%%%%%%%%%%%%%%%%%%%%%%%%%%%%%%%%%%%%%%%%%%%%%%

\section{Acknowledgement}
\label{Sec:6}
This work was partially supported by the grant no. DEC-2016/23/B/ST9/03114 from the Polish National
Science Center. We thank the anonymous referee for his/her careful reading of the manuscript and useful comments and suggestions. We are grateful to Andrzej Zdziarski and Włodek Kluzniak for their insightful comments and discussions. We also thank Eleonora Veronica Lai and Swayamtrupta Panda for some helpful discussions.  We acknowledge the support from the Interdisciplinary Center for Mathematical Modeling of the Warsaw University, through the computational grant Gb79-9, and the PL-Grid computational resources through the grant grb3.

%%%%%%%%%%%%%%%%%%%%%%%%%%%%%%%%%%%%%%%%%%%%%%%%%%%%%%%%%%%%%%%%%%%%%%%%%%%%%%%%%%%%%%%%%%%%%%%%%%%%%

\bibliography{ishika}{}

\begin{thebibliography}{}
\expandafter\ifx\csname natexlab\endcsname\relax\def\natexlab#1{#1}\fi
\providecommand{\url}[1]{\href{#1}{#1}}
\providecommand{\dodoi}[1]{doi:~\href{http://doi.org/#1}{\nolinkurl{#1}}}
\providecommand{\doeprint}[1]{\href{http://ascl.net/#1}{\nolinkurl{http://ascl.net/#1}}}
\providecommand{\doarXiv}[1]{\href{https://arxiv.org/abs/#1}{\nolinkurl{https://arxiv.org/abs/#1}}}

\bibitem[{Abramowicz \& Zurek(1981)}]{abramowicz1981rotation}
Abramowicz, M., \& Zurek, W. 1981, The Astrophysical Journal, 246, 314

\bibitem[{Acharya {et~al.}(2002)Acharya, Chakrabarti, \&
  Molteni}]{acharya2002interaction}
Acharya, K., Chakrabarti, S.~K., \& Molteni, D. 2002, Journal of Astrophysics
  and Astronomy, 23, 155

\bibitem[{Ahmad {et~al.}(2018)Ahmad, Misra, Iqbal, Maqbool, \&
  Hamid}]{ahmad2018modeling}
Ahmad, N., Misra, R., Iqbal, N., Maqbool, B., \& Hamid, M. 2018, New Astronomy,
  58, 84

\bibitem[{Axelsson \& Done(2018)}]{axelsson2018breaking}
Axelsson, M., \& Done, C. 2018, Monthly Notices of the Royal Astronomical
  Society, 480, 751

\bibitem[{Ba{\l}uci{\'n}ska-Church {et~al.}(2000)Ba{\l}uci{\'n}ska-Church,
  Church, Charles, Nagase, LaSala, \& Barnard}]{balucinska2000distribution}
Ba{\l}uci{\'n}ska-Church, M., Church, M., Charles, P., {et~al.} 2000, Monthly
  Notices of the Royal Astronomical Society, 311, 861

\bibitem[{Basak {et~al.}(2017)Basak, Zdziarski, Parker, \&
  Islam}]{basak2017analysis}
Basak, R., Zdziarski, A.~A., Parker, M., \& Islam, N. 2017, Monthly Notices of
  the Royal Astronomical Society, 472, 4220

\bibitem[{Cabanac {et~al.}(2010)Cabanac, Henri, Petrucci, Malzac, Ferreira, \&
  Belloni}]{cabanac2010variability}
Cabanac, C., Henri, G., Petrucci, P.-O., {et~al.} 2010, Monthly Notices of the
  Royal Astronomical Society, 404, 738

\bibitem[{Castor {et~al.}(1975)Castor, Abbott, \& Klein}]{castor1975radiation}
Castor, J.~I., Abbott, D.~C., \& Klein, R.~I. 1975, The Astrophysical Journal,
  195, 157

\bibitem[{Chakrabarti(1996)}]{Chakrabarti:1996ns}
Chakrabarti, S.~K. 1996, Astrophys. J., 464, 664, \dodoi{10.1086/177354}

\bibitem[{Chakrabarti(1997{\natexlab{a}})}]{chakrabarti1997spectral}
---. 1997{\natexlab{a}}, The Astrophysical Journal, 484, 313

\bibitem[{Chakrabarti(1997{\natexlab{b}})}]{Chakrabarti:1997hs}
---. 1997{\natexlab{b}}, Astrophys. J., 484, 313, \dodoi{10.1086/304325}

\bibitem[{Chakrabarti \& Titarchuk(1995{\natexlab{a}})}]{Chakrabarti:1995mx}
Chakrabarti, S.~K., \& Titarchuk, L.~G. 1995{\natexlab{a}}, Astrophys. J., 455,
  623, \dodoi{10.1086/176610}

\bibitem[{Chakrabarti \&
  Titarchuk(1995{\natexlab{b}})}]{chakrabarti1995spectral}
---. 1995{\natexlab{b}}, arXiv preprint astro-ph/9510005

\bibitem[{Chen {et~al.}(1995)Chen, Abramowicz, Lasota, Narayan, \&
  Yi}]{Chen:1995uc}
Chen, X.-m., Abramowicz, M.~A., Lasota, J.-P., Narayan, R., \& Yi, I. 1995,
  Astrophys. J. Lett., 443, L61, \dodoi{10.1086/187836}

\bibitem[{Cui {et~al.}(1997)Cui, Zhang, Focke, \& Swank}]{cui1997temporal}
Cui, W., Zhang, S., Focke, W., \& Swank, J. 1997, The Astrophysical Journal,
  484, 383

\bibitem[{Das(2002)}]{das2002generalized}
Das, T.~K. 2002, The Astrophysical Journal, 577, 880

\bibitem[{Das \& Chakrabarti(1999)}]{das1999mass}
Das, T.~K., \& Chakrabarti, S.~K. 1999, Classical and Quantum Gravity, 16, 3879

\bibitem[{{Das} \& {Czerny}(2012)}]{2012NewA...17..254D}
{Das}, T.~K., \& {Czerny}, B. 2012, \na, 17, 254

\bibitem[{{Das} {et~al.}(2003){Das}, {Pendharkar}, \& {Mitra}}]{das2003}
{Das}, T.~K., {Pendharkar}, J.~K., \& {Mitra}, S. 2003, \apj, 592, 1078,
  \dodoi{10.1086/375732}

\bibitem[{Done {et~al.}(2007)Done, Gierli{\'n}ski, \&
  Kubota}]{done2007modelling}
Done, C., Gierli{\'n}ski, M., \& Kubota, A. 2007, The Astronomy and
  Astrophysics Review, 15, 1

\bibitem[{Dubus {et~al.}(2019)Dubus, Done, Tetarenko, \&
  Hameury}]{dubus2019impact}
Dubus, G., Done, C., Tetarenko, B.~E., \& Hameury, J.-M. 2019, Astronomy \&
  Astrophysics, 632, A40

\bibitem[{{Fabian} {et~al.}(2015){Fabian}, {Lohfink}, {Kara}, {Parker},
  {Vasudevan}, \& {Reynolds}}]{fabian2015}
{Fabian}, A.~C., {Lohfink}, A., {Kara}, E., {et~al.} 2015, \mnras, 451, 4375,
  \dodoi{10.1093/mnras/stv1218}

\bibitem[{Feldmeier {et~al.}(1997)Feldmeier, Puls, \&
  Pauldrach}]{feldmeier1997possible}
Feldmeier, A., Puls, J., \& Pauldrach, A. 1997, Astronomy and Astrophysics,
  322, 878

\bibitem[{Fullerton {et~al.}(2006)Fullerton, Massa, \&
  Prinja}]{fullerton2006discordance}
Fullerton, A.~W., Massa, D., \& Prinja, R. 2006, The Astrophysical Journal,
  637, 1025

\bibitem[{Gammie {et~al.}(2003)Gammie, McKinney, \& T{\'o}th}]{gammie2003harm}
Gammie, C.~F., McKinney, J.~C., \& T{\'o}th, G. 2003, The Astrophysical
  Journal, 589, 444

\bibitem[{Grinberg {et~al.}(2014)Grinberg, Pottschmidt, B{\"o}ck, Schmid,
  Nowak, Uttley, Tomsick, Rodriguez, Hell, Markowitz,
  {et~al.}}]{grinberg2014long}
Grinberg, V., Pottschmidt, K., B{\"o}ck, M., {et~al.} 2014, Astronomy \&
  Astrophysics, 565, A1

\bibitem[{Hanke {et~al.}(2008)Hanke, Wilms, Nowak, Pottschmidt, Schulz, \&
  Lee}]{hanke2008chandra}
Hanke, M., Wilms, J., Nowak, M.~A., {et~al.} 2008, The Astrophysical Journal,
  690, 330

\bibitem[{Herrero {et~al.}(1995)Herrero, Kudritzki, Gabler, Vilchez, \&
  Gabler}]{herrero1995fundamental}
Herrero, A., Kudritzki, R., Gabler, R., Vilchez, J., \& Gabler, A. 1995,
  Astronomy and Astrophysics, 297, 556

\bibitem[{Hirsch {et~al.}(2019)Hirsch, Hell, Grinberg, Ballhausen, Nowak,
  Pottschmidt, Schulz, Dauser, Hanke, Kallman, {et~al.}}]{hirsch2019chandra}
Hirsch, M., Hell, N., Grinberg, V., {et~al.} 2019, Astronomy \& Astrophysics,
  626, A64

\bibitem[{Ichimaru(1977)}]{ichimaru1977bimodal}
Ichimaru, S. 1977, The Astrophysical Journal, 214, 840

\bibitem[{{Ingram} \& {Done}(2012)}]{ingram2012}
{Ingram}, A., \& {Done}, C. 2012, \mnras, 419, 2369,
  \dodoi{10.1111/j.1365-2966.2011.19885.x}

\bibitem[{Ingram \& Klis(2013)}]{ingram2013exact}
Ingram, A., \& Klis, M. v.~d. 2013, Monthly Notices of the Royal Astronomical
  Society, 434, 1476

\bibitem[{Janiuk {et~al.}(2018)Janiuk, Sapountzis, Mortier, \&
  Janiuk}]{JSFI177}
Janiuk, A., Sapountzis, K., Mortier, J., \& Janiuk, I. 2018, Supercomputing
  Frontiers and Innovations, 5

\bibitem[{{Jiang} {et~al.}(2019){Jiang}, {Blaes}, {Stone}, \&
  {Davis}}]{Jiang2019}
{Jiang}, Y.-F., {Blaes}, O., {Stone}, J.~M., \& {Davis}, S.~W. 2019, \apj, 885,
  144, \dodoi{10.3847/1538-4357/ab4a00}

\bibitem[{Kara {et~al.}(2019)Kara, Steiner, Fabian, Cackett, Uttley, Remillard,
  Gendreau, Arzoumanian, Altamirano, Eikenberry, {et~al.}}]{kara2019corona}
Kara, E., Steiner, J., Fabian, A., {et~al.} 2019, Nature, 565, 198

\bibitem[{Kawano {et~al.}(2017)Kawano, Done, Yamada, Takahashi, Axelsson, \&
  Fukazawa}]{kawano2017black}
Kawano, T., Done, C., Yamada, S., {et~al.} 2017, Publications of the
  Astronomical Society of Japan, 69, 36

\bibitem[{{Kotov} {et~al.}(2001){Kotov}, {Churazov}, \& {Gilfanov}}]{kotov2001}
{Kotov}, O., {Churazov}, E., \& {Gilfanov}, M. 2001, \mnras, 327, 799,
  \dodoi{10.1046/j.1365-8711.2001.04769.x}

\bibitem[{Kumar \& Chattopadhyay(2014)}]{Kumar:2014iwa}
Kumar, R., \& Chattopadhyay, I. 2014, Mon. Not. Roy. Astron. Soc., 443, 3444,
  \dodoi{10.1093/mnras/stu1389}

\bibitem[{Lewin {et~al.}(1997)Lewin, van~den Heuvel, \& van
  Paradijs}]{lewin1997x}
Lewin, W.~H., van~den Heuvel, E.~P., \& van Paradijs, J. 1997, X-ray Binaries,
  Vol.~26 (Cambridge University Press)

\bibitem[{Liu {et~al.}(2006)Liu, Van~Paradijs, \& Van
  Den~Heuvel}]{liu2006catalogue}
Liu, Q., Van~Paradijs, J., \& Van Den~Heuvel, E. 2006, Astronomy \&
  Astrophysics, 455, 1165

\bibitem[{Lucy \& Solomon(1970)}]{lucy1970mass}
Lucy, L., \& Solomon, P. 1970, The Astrophysical Journal, 159, 879

\bibitem[{{Lyubarskii}(1997)}]{lyubarskii1997}
{Lyubarskii}, Y.~E. 1997, \mnras, 292, 679, \dodoi{10.1093/mnras/292.3.679}

\bibitem[{Mandal \& Chakrabarti(2007)}]{mandal2007spectral}
Mandal, S., \& Chakrabarti, S.~K. 2007, Astrophysics and Space Science, 309,
  305

\bibitem[{Martins {et~al.}(2015)Martins, Boissier, Buat, Cambr{\'e}sy,
  {et~al.}}]{martins2015mass}
Martins, F., Boissier, S., Buat, V., Cambr{\'e}sy, L., {et~al.} 2015, sf2a, 343

\bibitem[{Mastroserio {et~al.}(2019)Mastroserio, Ingram, \& van~der
  Klis}]{mastroserio2019x}
Mastroserio, G., Ingram, A., \& van~der Klis, M. 2019, Monthly Notices of the
  Royal Astronomical Society, 488, 348

\bibitem[{{Meyer-Hofmeister} {et~al.}(2020){Meyer-Hofmeister}, {Liu}, {Qiao},
  \& {Taam}}]{2020arXiv200413241M}
{Meyer-Hofmeister}, E., {Liu}, B.~F., {Qiao}, E., \& {Taam}, R.~E. 2020, arXiv
  e-prints, arXiv:2004.13241

\bibitem[{Mineo {et~al.}(2012)Mineo, Gilfanov, \& Sunyaev}]{mineo2012x}
Mineo, S., Gilfanov, M., \& Sunyaev, R. 2012, Monthly Notices of the Royal
  Astronomical Society, 419, 2095

\bibitem[{Miškovičová {et~al.}(2016)Miškovičová, Hell, Hanke, Nowak,
  Pottschmidt, Schulz, Grinberg, Duro, Madej, Lohfink, \&
  et~al.}]{Mi2016chandra}
Miškovičová, I., Hell, N., Hanke, M., {et~al.} 2016, Astronomy \&
  Astrophysics, 590, A114

\bibitem[{Murdin \& Webster(1971)}]{murdin1971optical}
Murdin, P., \& Webster, B.~L. 1971, Nature, 233, 110

\bibitem[{Narayan \& Yi(1994)}]{Narayan:1994xi}
Narayan, R., \& Yi, I.-s. 1994, Astrophys. J. Lett., 428, L13,
  \dodoi{10.1086/187381}

\bibitem[{Oskinova {et~al.}(2012)Oskinova, Feldmeier, \&
  Kretschmar}]{oskinova2012clumped}
Oskinova, L.~M., Feldmeier, A., \& Kretschmar, P. 2012, Monthly Notices of the
  Royal Astronomical Society, 421, 2820

\bibitem[{Owocki {et~al.}(1988)Owocki, Castor, \& Rybicki}]{owocki1988time}
Owocki, S.~P., Castor, J.~I., \& Rybicki, G.~B. 1988, The Astrophysical
  Journal, 335, 914

\bibitem[{Paczynski(1977)}]{paczynski1977model}
Paczynski, B. 1977, The Astrophysical Journal, 216, 822

\bibitem[{Palit {et~al.}(2019)Palit, Janiuk, \& Sukova}]{palit2019effects}
Palit, I., Janiuk, A., \& Sukova, P. 2019, Monthly Notices of the Royal
  Astronomical Society, 487, 755

\bibitem[{Pooley {et~al.}(1999)Pooley, Fender, \&
  Brocksopp}]{pooley1999orbital}
Pooley, G.~G., Fender, R.~P., \& Brocksopp, C. 1999, Monthly Notices of the
  Royal Astronomical Society, 302, L1

\bibitem[{{Popham} \& {Sunyaev}(2001)}]{popham2001}
{Popham}, R., \& {Sunyaev}, R. 2001, \apj, 547, 355, \dodoi{10.1086/318336}

\bibitem[{Pottschmidt {et~al.}(2003)Pottschmidt, Wilms, Nowak, Pooley,
  Gleissner, Heindl, Smith, Remillard, \& Staubert}]{pottschmidt2003long}
Pottschmidt, K., Wilms, J., Nowak, M., {et~al.} 2003, Astronomy \&
  Astrophysics, 407, 1039

\bibitem[{Puls {et~al.}(2008)Puls, Markova, \& Scuderi}]{Puls:2006fk}
Puls, J., Markova, N., \& Scuderi, S. 2008, ASP Conf. Ser., 388, 101.
\newblock \doarXiv{astro-ph/0607290}

\bibitem[{Puls {et~al.}(2006)Puls, Markova, Scuderi, Stanghellini, Taranova,
  Burnley, \& Howarth}]{puls2006bright}
Puls, J., Markova, N., Scuderi, S., {et~al.} 2006, Astronomy \& Astrophysics,
  454, 625

\bibitem[{Rees {et~al.}(1982)Rees, Phinney, Begelman, \&
  Blandford}]{Rees:1982pe}
Rees, M., Phinney, E., Begelman, M., \& Blandford, R. 1982, Nature, 295, 17,
  \dodoi{10.1038/295017a0}

\bibitem[{Remillard \& McClintock(2006)}]{remillard2006x}
Remillard, R.~A., \& McClintock, J.~E. 2006, Annu. Rev. Astron. Astrophys., 44,
  49

\bibitem[{Ritter \& Kolb(2003)}]{ritter2003catalogue}
Ritter, H., \& Kolb, U. 2003, Astronomy \& Astrophysics, 404, 301

\bibitem[{Romero {et~al.}(2002)Romero, Bernad{\'o}, \&
  Mirabel}]{romero2002recurrent}
Romero, G.~E., Bernad{\'o}, M.~K., \& Mirabel, I. 2002, Astronomy \&
  Astrophysics, 393, L61

\bibitem[{{R{\'o}{\.z}a{\'n}ska} \& {Czerny}(2000)}]{rozanska2000}
{R{\'o}{\.z}a{\'n}ska}, A., \& {Czerny}, B. 2000, \aap, 360, 1170.
\newblock \doarXiv{astro-ph/0004158}

\bibitem[{Sukov{\'a} {et~al.}(2017)Sukov{\'a}, Charzy{\'n}ski, \&
  Janiuk}]{sukova2017shocks}
Sukov{\'a}, P., Charzy{\'n}ski, S., \& Janiuk, A. 2017, Monthly Notices of the
  Royal Astronomical Society, 472, 4327

\bibitem[{Sukov{\'a} \& Janiuk(2015)}]{sukova2015oscillating}
Sukov{\'a}, P., \& Janiuk, A. 2015, Monthly Notices of the Royal Astronomical
  Society, 447, 1565

\bibitem[{Sundqvist {et~al.}(2018)Sundqvist, Owocki, \& Puls}]{sundqvist20182d}
Sundqvist, J., Owocki, S., \& Puls, J. 2018, Astronomy \& Astrophysics, 611,
  A17

\bibitem[{Sundqvist \& Owocki(2013)}]{sundqvist2013clumping}
Sundqvist, J.~O., \& Owocki, S.~P. 2013, Monthly Notices of the Royal
  Astronomical Society, 428, 1837

\bibitem[{{Sunyaev} \& {Revnivtsev}(2000)}]{2000A&A...358..617S}
{Sunyaev}, R., \& {Revnivtsev}, M. 2000, \aap, 358, 617

\bibitem[{Taam \& Bodenheimer(1989)}]{taam1989double}
Taam, R.~E., \& Bodenheimer, P. 1989, The Astrophysical Journal, 337, 849

\bibitem[{Taam \& Sandquist(2000)}]{taam2000common}
Taam, R.~E., \& Sandquist, E.~L. 2000, Annual Review of Astronomy and
  Astrophysics, 38, 113

\bibitem[{Tananbaum {et~al.}(1972)Tananbaum, Gursky, Kellogg, Giacconi, \&
  Jones}]{tananbaum1972observation}
Tananbaum, H., Gursky, H., Kellogg, E., Giacconi, R., \& Jones, C. 1972, The
  Astrophysical Journal, 177, L5

\bibitem[{{{\v{C}}echura} \& {Hadrava}(2015)}]{2015A&A...575A...5C}
{{\v{C}}echura}, J., \& {Hadrava}, P. 2015, \aap, 575, A5

\bibitem[{Zdziarski \& De~Marco(2020)}]{zdziarski2020two}
Zdziarski, A.~A., \& De~Marco, B. 2020, arXiv preprint arXiv:2002.04652

\bibitem[{Zhang {et~al.}(1997)Zhang, Cui, Harmon, Paciesas, Remillard, \&
  Van~Paradijs}]{zhang19971996}
Zhang, S., Cui, W., Harmon, B., {et~al.} 1997, The Astrophysical Journal
  Letters, 477, L95

\bibitem[{Zi{\'o}{\l}kowski(2005)}]{ziolkowski2005evolutionary}
Zi{\'o}{\l}kowski, J. 2005, Monthly Notices of the Royal Astronomical Society,
  358, 851

\bibitem[{Zi{\'o}{\l}kowski(2014)}]{ziolkowski2014determination}
---. 2014, Monthly Notices of the Royal Astronomical Society: Letters, 440, L61

\end{thebibliography}
\bibliographystyle{aasjournal}

\end{document}